\documentclass{PoS}

\usepackage{graphicx}
\graphicspath{{./Figs/}}
\usepackage{subcaption}
\usepackage{amsmath}
\usepackage{slashed}
\usepackage[sort&compress,numbers,merge]{natbib}
\usepackage{natbib,natbibspacing}

\newcommand{\Tr}{\operatorname{Tr}}

\renewcommand{\Im}{\mathfrak{Im}}

\newcommand{\Fig}[1]{\mbox{{Fig.\,\ref{#1}}}}
\newcommand{\Eq}[1]{\mbox{Eq.\,\eqref{#1}}}

\title{Triple-gluon and quark-gluon vertex from lattice QCD in Landau gauge}

\ShortTitle{Triple-gluon and quark-gluon vertex from lattice QCD in Landau gauge}

\author{\speaker{A.\ Sternbeck}$^1$, P.-H.\ Balduf$^{1,2}$, 
 A.\ K{\i}z{\i}lers\"u$^3$, O.\ Oliveira$^4$, P.J. Silva$^4$, 
 J.\ Skullerud$^{5}$, \hbox{A.\ G.\ Williams}$^{3}$\\

$^1$ Theoretisch-Physikalisches Institut, Friedrich-Schiller-Universit\"at Jena, 07743 Jena, Germany\\
$^2$ Institut f\"ur Physik, Humboldt-Universit\"at zu Berlin, 12489 Berlin, Germany\\
$^3$ CSSM, Department of Physics, University of Adelaide, Adelaide, SA 5005, Australia\\
$^4$ CFisUC, Department of Physics, University of Coimbra, P--3004 516 Coimbra, Portugal \\ 
$^5$ Department of Mathematical Physics, National University of Ireland Maynooth, Ireland\\

\vspace{-2ex}E-mail: \email{andre.sternbeck@uni-jena.de}
}

\abstract{We report on preliminary results for the triple-gluon and the 
quark-gluon vertex in Landau gauge. Our results are based on two-flavor and 
quenched lattice QCD calculations for different quark masses, lattice spacings 
and volumes. We discuss the momentum dependence of some of the vertices's form
factors and the deviations from the tree-level form.}

\FullConference{34th annual International Symposium on Lattice Field Theory\\
		24-30 July 2016\\  University of Southampton, UK}

\begin{document}

\section{Motivation}

Lattice QCD calculations are currently the prime tool for theoretical 
model-free studies of hadronic properties (or QCD in general) beyond the 
regime of perturbation theory. There are however also other nonperturbative 
methods which may provide a look at strong interaction physics that complements the lattice. An example 
are the bound-state and Dyson-Schwinger equations (DSEs) of QCD  (see 
\cite{Eichmann:2016yit} for a recent review). Like the lattice, this approach 
starts from first principles, namely the partition function, but in addition it 
requires fixing a gauge. In principle, any physics extracted with this method is independent 
of the gauge, for example, masses or form factors via correspondingly 
defined bound-state equations. The numerical treatment, however, requires 
a truncation of the infinite tower of DSEs and so a truncation 
(resp.\ gauge) dependence may enter. 

Let us elucidate this with an example: In a relativistic quantum field theory 
hadronic states appear as poles in the spectral decomposition of $n$-point 
Green's functions. The residues of these poles ($P^2=-M^2$) define the wave 
function of the corresponding bound state, which in the case of mesons, 
satisfy the Bethe-Salpether (BS) equation. This equation is typically written
for the amputated wave function, aka Bethe-Salpether amplitude, and describes
a meson as two-particle system of a strongly interacting quark and 
anti-quark. It reads \cite{Eichmann:2016yit}
\begin{equation}
 \Gamma_{\alpha\beta}(P,p) =  \int \frac{d^4q}{(2\pi)^4} K_{\alpha\gamma,\delta\beta}(p,q,P)
   \big\{S(q_+)\, \Gamma(P,q)\, S(q_-)\big\}_{\gamma\delta}
   \label{eq:BSE}
\end{equation}
where $\Gamma(p,P)$ is the BS amplitude which contains all the properties 
of the meson. $S(q_{+})$ and $S(q_{-})$ are the respective quark 
propagators and the scattering kernel $K(p,q,P)$ encodes all possible 
(two-particle-irreducible) interactions between them. $P$ is the total and $p$ 
the relative momentum and $q_{\pm}=q\pm P/2$. If $S$ and $K$ were known for all momenta, a solution of 
\Eq{eq:BSE} would yield $\Gamma(p,P)$ from which masses, form factors and the 
like can be extracted. In fact, the masses of ground and excited states are
given by all $P_i=-M_i^2$ for which $\Gamma(q,P_i)$ is a solution to \Eq{eq:BSE}.

In practice, however, neither $S$ nor $K$ are fully available. $S$ satisfies a DSE
which itself needs the gluon propagator and the quark-gluon vertex as input. 
Both satisfy their own DSEs which involve other $n$-point functions. 
A truncation of the DSEs and the scattering kernel $K$ is thus needed. 

A popular truncation is the \emph{rainbow-ladder} (RL) truncation. It
substitutes the gluon propagator by an effective propagator and reduces the
quark-gluon vertex to its tree-level form. Also the scattering kernel $K$ 
is reduced to one (effective) gluon exchange. This often is sufficient for
ground states which are found to be mostly only sensitive to the integrated 
scattering kernel, but observables such as excited states typically require a 
beyond rainbow-ladder treatment and this is where lattice results of the quark-gluon
vertex can provide important input\footnote{I thank G.\ Eichmann for briefing me 
about the current status.}. Since the DSE of the quark-gluon vertex contains the 
triple-gluon vertex, lattice results for both provides additional information 
how to systematically improve the truncation of the quark-gluon-vertex DSE.

Such lattice calculations are demanding, because continuum extrapolated results 
are needed for the many different form factors of the vertex functions. To this 
end, one needs good control over finite volume and lattice discretization
effects. In particular the latter can be challenging due to the notorious 
hypercubic lattice artifacts. 

Here we present the first steps of a long-term project. Our calculations were 
performed on the $N_f=2$ gauge field ensembles of the 
RQCD collaboration and, in addition, on three quenched ensembles provided by us. 
Lattice spacings and volumes were chosen such that a comparison with the 
unquenched data is possible and that (for the triple-gluon vertex) low momenta
are reached (see Table~\ref{tab:stat_summary} for details). All ensembles were
fixed to Landau gauge before taking Monte-Carlo data for the required two- and 
three-point Green's functions.

\begin{table*} 
 \begin{minipage}[c]{0.22\textwidth}
   \caption{Summary of our $N_f=0,2$ gauge field ensembles. The
   physical parameters for the $N_f=2$ ensembles were taken from \cite{Bali:2014gha}.
   \vspace{4ex}}
 \label{tab:stat_summary}
 \end{minipage}\hfill
 \begin{minipage}[c]{0.72\textwidth}\small
   \begin{tabular}{l@{\qquad}c@{\quad}c@{\quad}c@{\quad\qquad}cc@{\quad}r}
   \hline\hline
  no. & $\beta$ & $\kappa$ & $L^3_s\times L_t$ & $a$ [fm] & $m_\pi$ [MeV] & 
\#configs.\\
   \hline
 III &  5.20 & 0.13596 & $32^3\times 64$       & 0.08 & 280 & 900 \\*[0.3ex]
 IV-a &  5.29 & 0.13620 & $32^3\times 64$       & 0.07 & 422 & 900 \\
 IV-b &  5.29 & 0.13632 & $32^3\times 64$       & 0.07 & 295 & 908 \\
 IV-c &  5.29 & 0.13632 & $64^3\times 64$       & 0.07 & 290 & 750 \\[0.3ex]
 V &  5.40 & 0.13647 & $32^3\times 64$       & 0.06 & 426 & 900 \\*[0.3ex]
   \hline
 IV-q &  6.16 & ---     & $32^3\times 64$       & 0.07 & --- & 1000 \\
 II-q & 5.70 & ---     & $48^3\times 96$       & 0.17 & --- & 1000 \\
 I-q & 5.60 & ---     & $72^3\times 72$       & 0.22 & --- & 699 \\
   \hline\hline
 \end{tabular}
 \end{minipage}
\end{table*}

\section{Lattice results for the triple-gluon vertex}

The Green's functions one needs for an analysis of the triple-gluon vertex are the gluon 
propagator $D^{ab}_{\mu\nu}(p)$ and the triple-gluon Green's function
\begin{equation}
  G^{abc}_{\mu\nu\rho}(p,q)=\left\langle A^a_\mu(p)\,A^b_\nu(q)\,A^c_\rho(k)\right\rangle_U
  \qquad\text{with}\quad k\equiv -(p+q)\,.
\end{equation}
To leading order in the lattice spacing, the gluon fields $A_\mu^a(p)$ are Fourier-transformed functions of the 
gauge-fixed link variables $U_{x\mu}$:
\begin{equation}
 A^a_\mu(p) = \frac{1}{\sqrt{V}}\sum_x e^{ipx} A_\mu^a(x) \quad\text{with}\quad 
 A^a_\mu(x) \equiv \frac{2}{ag_0}\,\Im\Tr T^a U_{x\mu}
\end{equation}
We have calculated these Green's function for the subset of (discrete) momenta with
$p^2=q^2$. To increase the signal-to-noise ratio we averaged our Monte-Carlo 
data over equal $\vert p\vert=\vert q\vert$ and also over nearby momenta before 
performing a time-series analysis. For the ''momentum smoothing`` we have 
tried different bin sizes 
$\Delta=\operatorname{abs}\left(\vert p_2\vert-\vert p_1\vert\right)=\pi/an$ 
with $n=128,64,32,16$ and found that already $a\Delta=\pi/128$ results in a significant improvement of the 
signal-to-noise ratio.

The tensor structure of the triple-gluon vertex is described by
14 form factors (see, e.g., \cite{Gracey:2011vw}). We are interested in the 
transverse part, which is accessible on the lattice and also is the 
interesting part in the Landau gauge. To the author's knowledge the full transverse 
part has never been addressed in a lattice study; only the projection of the 
vertex (where the color sum is implied) 
\begin{equation}
  G_1(p,q) = \frac{\Gamma^{(0)}_{\mu\nu\rho}}{\Gamma^{(0)}_{\mu\nu\rho}}
        \frac{G_{\mu\nu\rho}(p,q,p-q)}{D_{\mu\lambda}(p)
D_{\nu\sigma}(q)D_{\rho\omega}(p-q)\Gamma^{(0)}_{\lambda\sigma\omega}}
\end{equation}
on its tree-level form $\Gamma^{(0)}(p,q)$ has been analyzed for $SU(2)$ and 
$SU(3)$ Yang-Mills theories 
\cite{Cucchieri:2008qm,Athenodorou:2016oyh,Duarte:2016ieu}. Those studies we 
complement here with quenched and unquenched data for both $G_1$ and the 
transverse form factors. For the latter we use the Bose-symmetric form of
the transversely-projected triple-gluon vertex \cite{Eichmann:2014xya}
\begin{equation}
 \Gamma^T_{\mu\nu\rho}(p,q) 
   = \sum_{i=1}^{4}F_i(p^2,q^2,\phi)\,
\tau_{i}^{\mu\nu\rho}(p,q) 
\end{equation}
where the four transverse form factors $F_i$ are functions of $p^2$, $q^2$ and 
the angle $\phi$ between $p$ and $q$. The corresponding tensor base vectors 
$\tau_{i}$ can be found in \cite{Eichmann:2014xya}.

\begin{figure}[t]
 \centering
 \mbox{\includegraphics[width=0.49\textwidth]{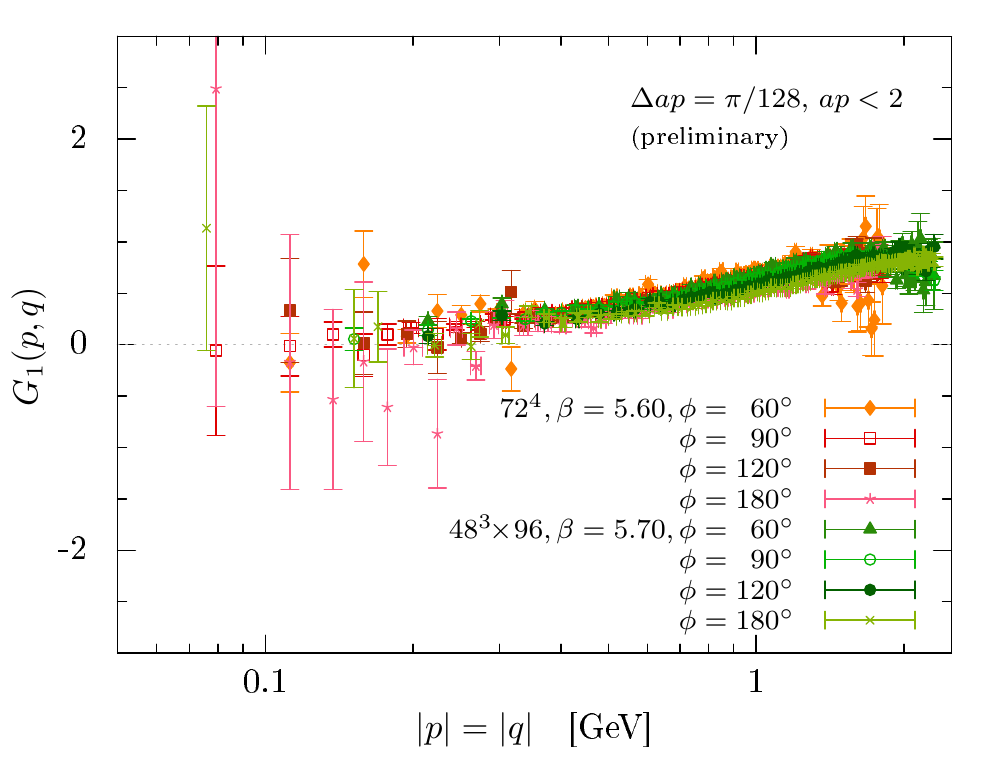}
 \quad
 \includegraphics[width=0.49\textwidth]{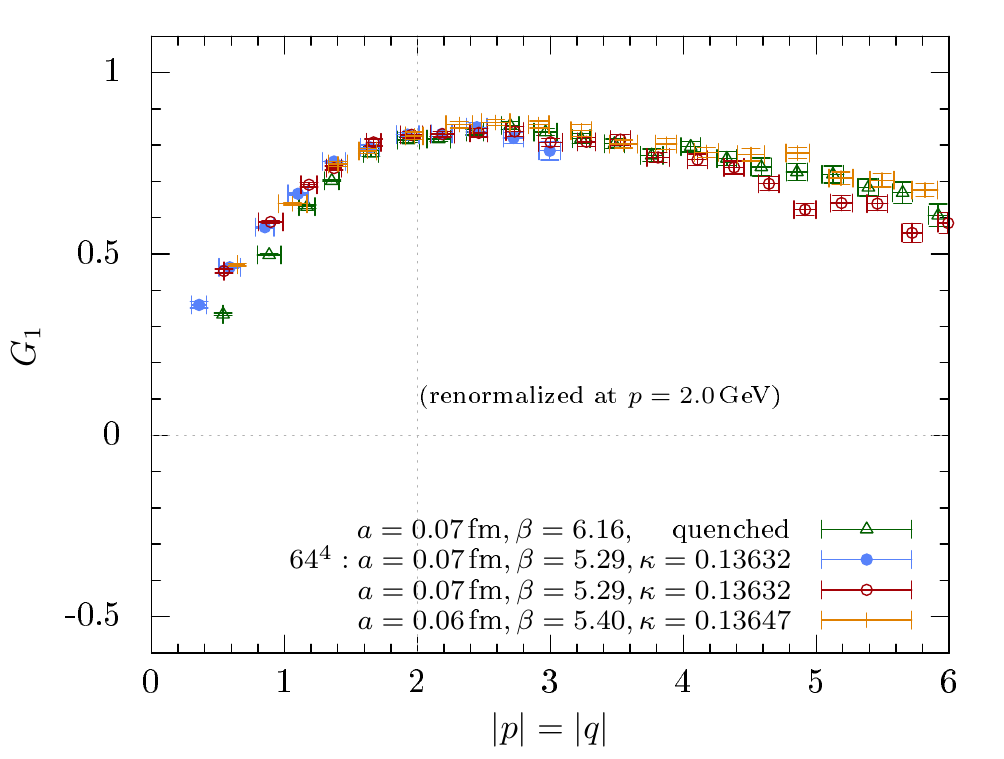}}
 \caption{$G_1(p,q)$ versus $\vert p\vert=\vert q\vert$. Left: quenched data
   at low momentum for different angles 
   $\phi=\sphericalangle(p,q)=60^\circ,90^\circ, 120^\circ$ and $180^\circ$. 
   Right: for a fixed angle $\phi=90^\circ$ but for different quark masses and 
   lattice spacings, and for a quenched calculation. Data for nearby-momenta has 
   been averaged. The bin size is $a\Delta=\pi/128$ in the left plot and 
   $a\Delta=\pi/32$ in the right.\vspace{-1ex}}
 \label{fig:projtau1_p}
\end{figure}

Data for $G_1$ versus $\vert p\vert = \vert q\vert$ is shown in 
\Fig{fig:projtau1_p}. The left panel shows quenched data at low momentum
for different angles $\phi$.  
The right panel compares quenched with unquenched data at a fixed angle $\phi=90^\circ$ 
but for different lattice spacings and quark masses. The momentum bin size is 
$a\Delta=\pi/128$ for the data on the left panel and $a\Delta=\pi/32$ on the 
right. To allow for a better comparison of the momentum dependence, the data
points in the right panel have been renormalized relatively to each other at 
$\mu=2\,\text{GeV}$. The data in the left panel were not renormalized.
There the focus is on the ''zero-crossing'' which was seen by others 
\cite{Cucchieri:2008qm,Athenodorou:2016oyh,Duarte:2016ieu}. In fact, two recent
lattice studies claim numerical evidence for $G_1$ crossing zero at about 
$\vert p\vert =100\ldots200\,\text{MeV}$ in \cite{Athenodorou:2016oyh} and 
$\vert p\vert =220\ldots260\,\text{MeV}$ in
\cite{Duarte:2016ieu}. 
Some of our data points between $\vert p\vert=100\ldots300\,\text{MeV}$ have 
negative values for $G_1$ (e.g., for $\phi=60^\circ$ and $\phi=180^\circ$), but 
there are also several points with positive $G_1$. With respect to the
rather large  error bars, at the moment we can confirm $G_1$ touches zero 
at about $\vert p\vert =100\ldots200\,\text{MeV}$, but there is no clear 
numerical evidence yet for a divergence at small momenta. Perhaps future studies 
at much smaller momenta will answer this. A more detailed discussion will 
follow in \cite{Balduf}.

Looking at the right panel of \Fig{fig:projtau1_p} we see quenching
effects on the triple-gluon vertex are rather small. With the renormalization 
point set at $\mu=2\,\text{GeV}$ we see the quenched data points below 
$\vert p\vert =2\,\text{GeV}$
fall off more rapidly towards smaller $\vert p\vert $ than the unquenched. 
This is consistent with what has been seen in a recent DSE study of this vertex
\cite{Williams:2015cvx} assuming that $F_1\approx G_1$. As mentioned, we also 
looked at the four transverse form factors $F_{i=1,\ldots,4}$. $F_4$ is zero in our 
kinematic, but the other three can be extracted. It turns out that $F_1$
by far is the dominant form factor, while $F_2$ and $F_3$ are rather small in 
comparison. The momentum dependence of $F_1$ thus roughly mirrors that
of $G_1$. A detailed analysis will follow \cite{Balduf}.

\section{Lattice results for the quark-gluon vertex}

The tensor structure of the quark-gluon vertex (in the continuum) consists of 12 
independent vectors. In analogy to the Ball-Chiu decomposition of its QED 
counterpart, it is often written as the sum of a term satisfying the Slavnov-Taylor 
identity (STI) and another term which is transverse to the 
gluon momentum $k$. That is, $\Gamma_\mu(p,k) = \Gamma^{ST}_\mu(p,k) + \Gamma^T_\mu(p,k)$
with $k_\mu\Gamma^T_\mu=0$ and
\begin{align}
 \Gamma^{ST}_\mu(p,k) &= \sum_{i=1\ldots 4} 
\lambda_i(p,k)\, L_{i\mu}(p,k) \quad\text{and}\quad
 \Gamma^T_\mu(p,k) = \sum_{i=1\ldots 8} \tau_i(p,k) \,
T_{i\mu}(p,k)\,.
\end{align}
We consider here only the form factors $\lambda_1$, $\lambda_2$ and $\lambda_3$. 
The corresponding base vectors read \cite{Skullerud:2002ge}
(all momenta are considered as outgoing)
\begin{equation}
 L_{1\mu}(p,k)=\gamma_\mu\,,
 \quad 
 L_{2\mu}(p,k)=-(2\slashed{p} + \slashed{k})(2p + k)_\mu
 \quad\text{and}\quad
 L_{3\mu}(p,k)=-i(2p + k)_\mu\,.
\end{equation}
$L_{4\mu}$ and the remaining (transverse) base vectors $T_{i\mu}$ can be found in 
\cite{Skullerud:2002ge}. The $T$'s are all zero for the \emph{soft-gluon} kinematic 
($k=0$) which we consider here throughout. In this kinematic also 
$\lambda_4(p,0)=0$ due to the STI and the other three $\lambda_i$ are functions of 
$p^2$ only, i.e., $\lambda_i(p,0)=\lambda_i(p^2,0)$.

On the lattice the tensor structure is actually even more complex due 
to the finite lattice spacing which breaks $O(4)$ invariance. For simplicity 
though, we will restrict ourselves to the continuum form, but consider lattice 
tree-level corrections to $L_{1\mu}$ (resp.\ $\lambda_1$). This will help us
to see where lattice discretisation effects are significant.

In QED, the $\lambda_i$'s for the fermion-photon vertex are functions of
the fermion propagator alone. This is due to the Ward-Takahashi identity. 
In QCD, however, the STI for the quark-gluon vertex also contains the ghost 
dressing function and the quark-ghost scattering kernel. A non-perturbative 
determination of the $\lambda_i$'s for the quark-gluon vertex thus requires a 
calculation of 2- and 3-point functions. These are the quark propagator $S(p)$, 
the gluon propagator $D_{\mu\nu}(k)$ and 
the quark--anti-quark--gluon 3-point function $V^{c, ij}_{\mu, \alpha\beta} 
= \left\langle \bar{\psi}^i_\alpha\psi^j_\beta A^c_\mu \right\rangle$ in 
momentum space. The form factors one gets then from the relation between
vertex and 2- and 3-point functions:
\begin{align*}
 V^{c, ij}_{\mu, \alpha\beta}(p,k) &= \Gamma^{d, rs}_{\lambda,\sigma\rho}(p,k) 
 \cdot S^{ir}_{\sigma\alpha}(p)\cdot D^{cd}_{\mu\lambda}(k)\cdot S^{js}_{\rho\beta}(q)
\quad
\text{with}\quad q = -(p+k)\,.
\end{align*}

We calculated $S$, $D$ and $V$ on all $32^3\times 64$ ensembles 
listed in Table \ref{tab:stat_summary}. For the calculation of the quark 
propagator we averaged over four point sources and to reduce off-shell $O(a)$
effects we used the ``rotated quark propagator`` of 
\cite{Skullerud:2000un,Skullerud:2001aw} with $b_q$ and $c_q$ set to their 
tree-level values. For the quenched ensembles we followed a partially quenched 
approach and set $\kappa=0.134$ and $c_{SW}=1.64$. According to 
\cite{Gockeler:1997fn} this should correspond to a pion mass value of 
$m_\pi\gtrsim 800\,\text{MeV}$.

\begin{figure*}
 \centering
 \mbox{\includegraphics[width=7.3cm]{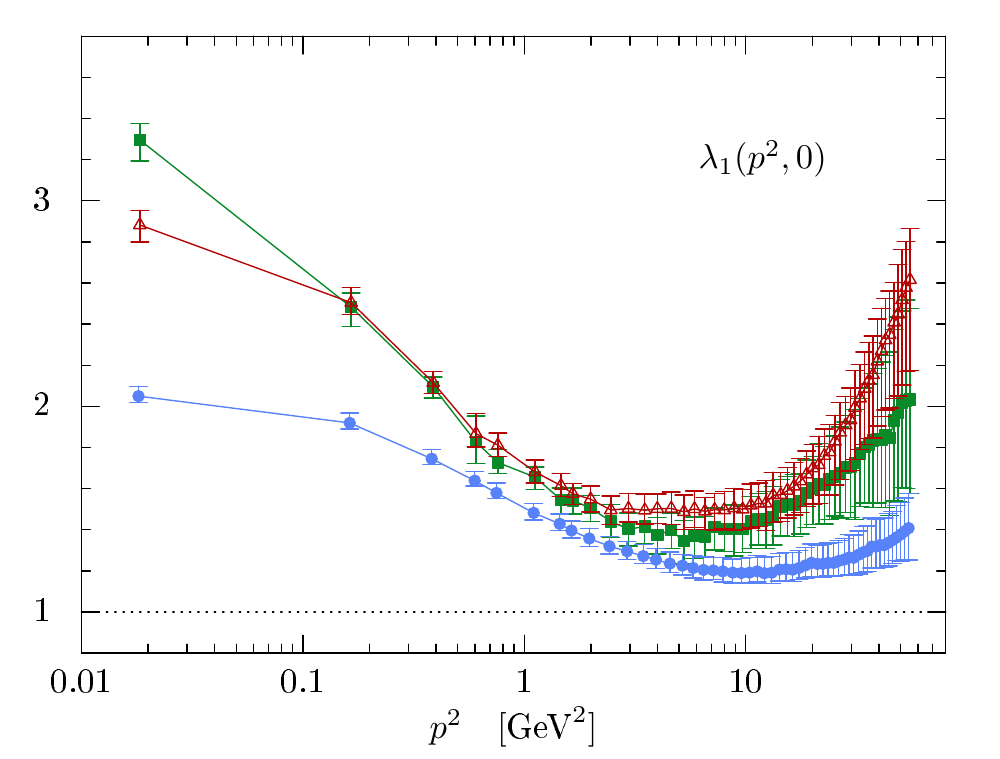}\quad
  \includegraphics[width=7.3cm]{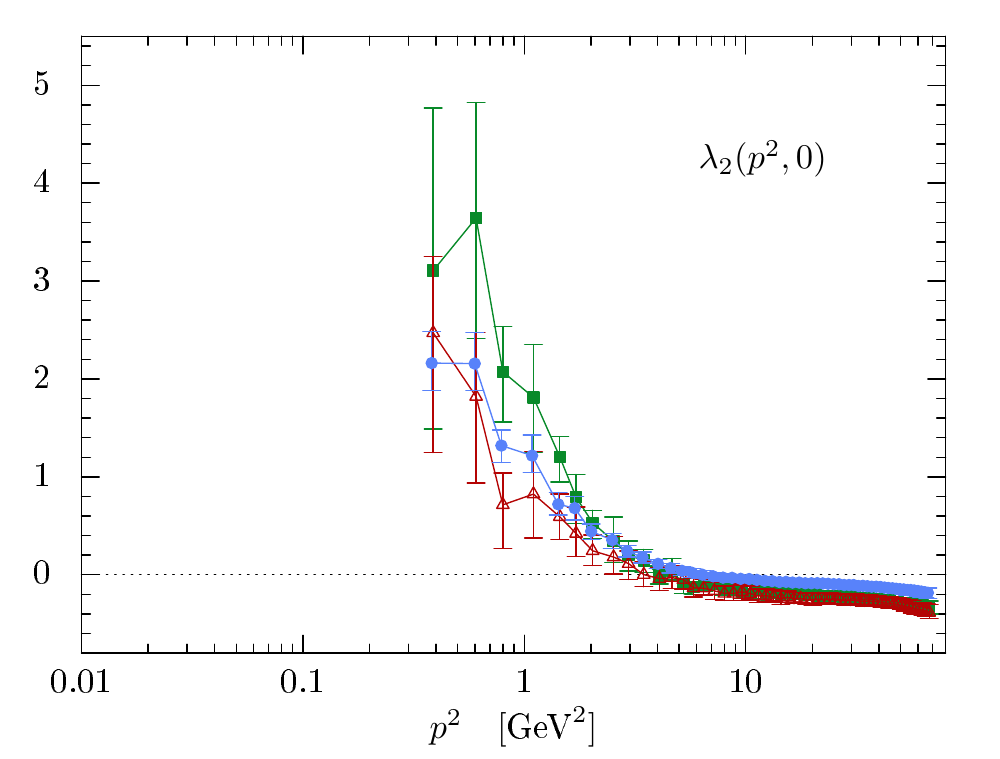}}\\
 \mbox{\includegraphics[width=7.3cm]{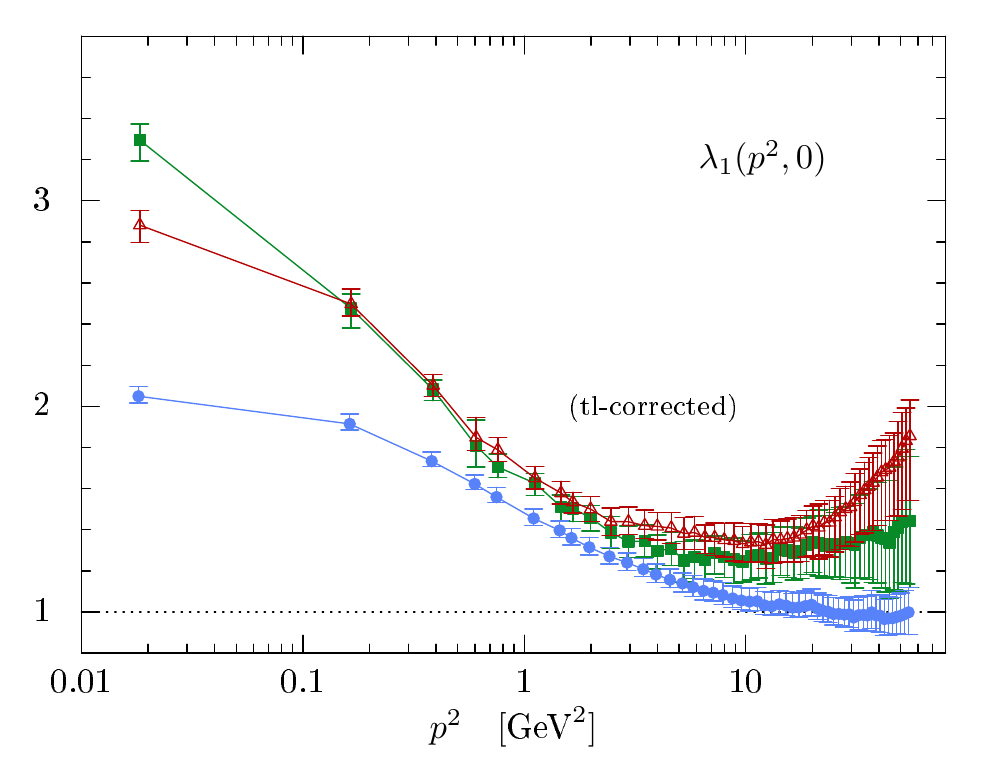}\quad
  \includegraphics[width=7.3cm]{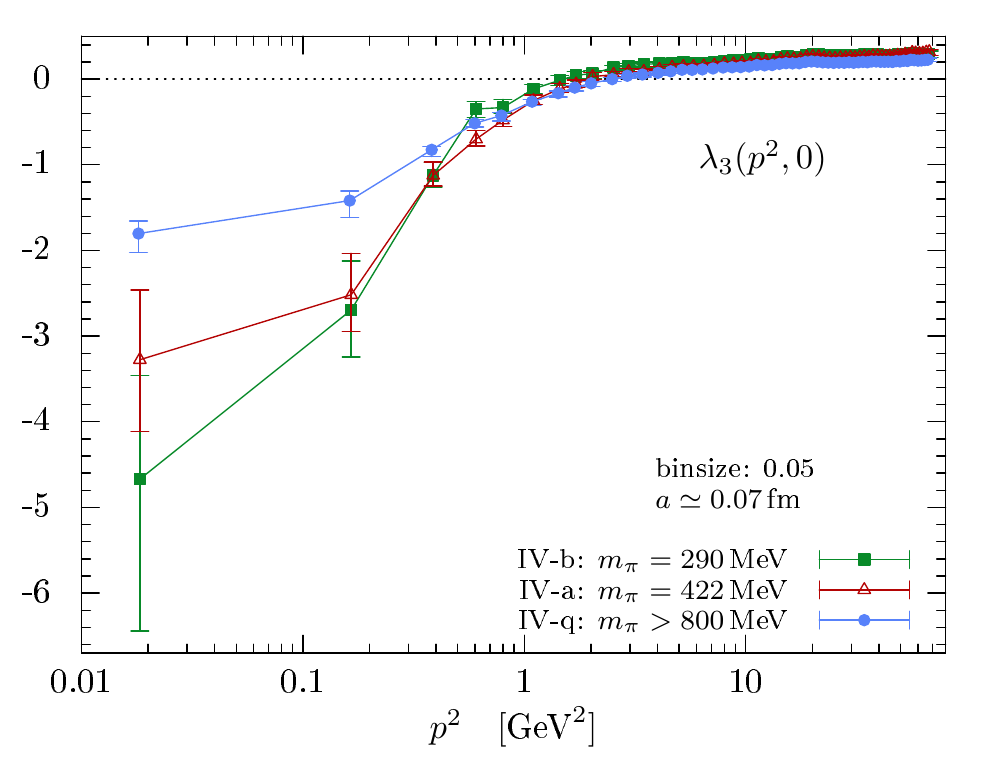}}
  \caption{
  $\lambda_1(p^2,0)$, $\lambda_2(p^2,0)$ and $\lambda_3(p^2,0)$ 
  from three calculations 
  at about the same lattice spacing ($a\simeq 0.07\,$fm) but for different 
  bare quark masses including a quenched simulation (IV-q). The legend
  in the bottom right panel applies to all plots. Shown are the unrenormalized 
  data. The bottom left panel show the corresponding
  lattice tree-level corrected data for $\lambda_1$.\vspace{-1ex}}
  \label{fig:lambda123_p2}
\end{figure*}

In \Fig{fig:lambda123_p2} we show some preliminary data for $\lambda_1(p^2,0)$, 
$\lambda_2(p^2,0)$ and $\lambda_3(p^2,0)$. The data are from three calculations with similar 
lattice spacings ($a\simeq0.07\,\text{fm}$) but different quark mass values. 
The corresponding ensembles are listed as IV-a, IV-b and IV-q in Table~\ref{tab:stat_summary}.
Our results from the lightest ensembles IV-b, that is for a pion mass of about 
$m_\pi=295\,\text{MeV}$ \cite{Bali:2014gha}, are the green squares in \Fig{fig:lambda123_p2}. 
Red triangles refer to the heavier unquenched
ensemble (IV-a, $m_\pi=422\,\text{MeV}$), while blue circles are for the
partially-quenched data (IV-q) with $m_\pi\gtrsim 800\,\text{MeV}$. The points have 
not been renormalized to visualize differences in the bare data. 

Looking first at the unquenched data (squares and triangles), we see that the 
quark mass effects for $\lambda_2$ and $\lambda_3$ are small for high momenta
but they grow towards lower $p^2$. For $\lambda_1$, on the other hand, the 
quark mass affects the behavior at high momentum, while at intermediate 
momentum the effect is small. Specifically, the two unquenched data sets for 
$\lambda_1$ match between $p^2=0.1\,\text{GeV}^2$ and $1\,\text{GeV}^2$, but 
below and above quark mass effects are seen.

Comparing the quenched and unquenched data sets for $\lambda_1$ and 
$\lambda_3$, we see the quenched sets have a slightly flatter momentum dependence
than the unquenched. We have checked that 
this cannot be compensated through a simple (multiplicative) renormalization. 
For $\lambda_2$, quenched and unquenched data more or less agree. There a 
renormalization of the quenched data relative to the unquenched points 
at large $a^2p^2$ would result in a stronger momentum dependence for small $p^2$. 

For $\lambda_1$ we also have lattice tree-level corrected results. By comparing
corrected to uncorrected results we see no difference in the data for 
$\lambda_1$ below $a^2p^2=1$. Above $a^2p^2=1$, however, there is a clear difference
and this difference also grows with $a^2p^2$. The tree-level correction brings 
the data closer to what is expected from perturbation theory. Our correction
is not perfect and leaves room for further improvements but it suggests the 
strong rise of $\lambda_1(p^2,0)$ towards larger $p^2$ is a lattice 
artifact. Note in \cite{Aguilar:2016lbe} it was argued $\lambda_1(p^2,0)$ 
should grow with $p^2$, but given the effect of the tree-level 
corrections one should be skeptical if our data will eventually confirm that.
It certainly deserves further study.

Our results for $\lambda_1$, $\lambda_2$ and $\lambda_3$ below $a^2p^2=1$, that is 
for momenta $p^2< 5\,\text{GeV}^2$, show no significant discretization
effects. Apart from volume effects, which we expect for our lowest momentum, 
we are thus confident our data resemble to a good approximation the correct 
non-perturbative behavior of $\lambda_1$, $\lambda_2$ and $\lambda_3$ between 
$p^2=0.1\ldots 5\,\text{GeV}^2$. In this regime we see that all three form 
factors show a strong deviation from the tree-level form below 
$p^2=1\,\text{GeV}^2$: $\lambda_1$ rises towards $p\to0$ but less strong than 
$\lambda_2$; $\lambda_3$ falls below zero towards $p\to0$. The value of the 
quark mass seems to have an effect on the slope; for $\lambda_2$ the effect is 
more pronounced than for $\lambda_1$ and $\lambda_3$. Above $p^2=2\,\text{GeV}^2$,
on the other hand, $\lambda_1$, $\lambda_2$ and $\lambda_3$, are rather close to 
their perturbative form. This finding may shed some light on the partial success and 
failure of the rainbow-ladder truncation, where only $\lambda_1$ is considered 
(modeled) but $\lambda_2$ and $\lambda_3$ are ignored.

\section{Summary}

We have presented first (preliminary) results for some of the many form factors 
of the triple-gluon and quark-gluon vertex of QCD in Landau gauge. The data
are the result of two separate projects \cite{Balduf,Kizilersu}
which however base on the same set of $N_f=2$ ensembles of the RQCD 
collaboration and a few additional quenched ensembles. We have touched some 
issues addressed in the recent literature and gave a first impression 
what the momentum behavior looks like. A detailed analysis will be reported 
in due course.

\bigskip
\centerline{------------------------------------------------}
\medskip
{\small 
Calculations were performed on the Cray system of the 
 North-German Supercomputing Alliance (\textsc{HLRN} project bep00046) and 
 the \textsc{Omega} cluster at the FSU Jena. We regret the sudden 
 death of M.\,M\"uller-Preu\ss{}ker who collaborated with us on this project. We thank the RQCD 
collaboration for giving us access to their $N_f=2$ gauge field configurations.
P.J.\,Silva was supported by FCT (grant SFRH/BPD/109971/2015).

\bibliography{references.bib}
\bibliographystyle{apsrev4-1}

}
\end{document}